\documentclass[11pt,twoside]{article}

\usepackage{asp2006}
\usepackage{epsf}
\usepackage{lscape}

\begin{document}
\title{Catching the Binaries Amongst B[e] Stars}   
\author{Michaela Kraus}   
\affil{Astronomick\'y \'ustav, Akademie v\v{e}d \v{C}esk\'e republiky, Fri\v{c}ova 298, 251\,65 Ond\v{r}ejov, Czech Republic}    
\author{Marcelo Borges Fernandes and Olivier Chesneau}   
\affil{UMR 6525 H. Fizeau, Univ. Nice Sophia Antipolis, CNRS, Observatoire de
la C\^{o}te d'Azur, Av. Copernic, F-06130 Grasse, France}

\begin{abstract} 
It is surprising to find dust around B type stars, as in the case of B[e] stars.
These stars exhibit a dense, dusty environment
witnessed by their infrared-excess and many emission lines from permitted
and forbidden transitions. Given the large uncertainties on their
distances, this spectral type gathers many different kind of sources that
may harbor a similar circumstellar environment, i.e. a dense dusty disk.
At the exception of Young Stellar Objects, in many cases, it is very
difficult to understand the origin of such a disk without invoking
binarity.
We describe current powerful methods, like spectral disentangling, spectro-astrometry 
and long baseline interferometry, to detect especially close binaries amongst the
unclassified B[e] stars. The role of binary mergers in the formation of the B[e] 
phenomenon, especially in supergiants and compact PNe, is also discussed. 
\end{abstract}


\section{Introduction}   

B[e] stars are stars of spectral type B with plenty of emission lines from permitted
and especially forbidden transitions in predominantely neutral or low-ionized metals 
like O{\sc i} and Fe{\sc ii}. In addition, these stars show strong infrared excess
emission due to warm and hot circumstellar dust. Since these spectral characteristics
are related purely to the circumstellar medium of these objects, it is obvious that 
they might be found in stars in quite different evolutionary phases. And indeed, a 
detailed investigation of the B[e] stars performed by \citet{MK_Lamers} revealed that 
B[e] stars can be pre-main sequence as well as post-main sequence in nature (see 
Table\,\ref{MK_classes}).

\begin{table}[!ht]
\caption{B[e] phenomenon in various types of stars. Listed are the different B[e]
characteristics and their origin. Only the symbiotics are known to be interacting
binaries, while in the other B[e] classes a binary component is not needed (but see
Sect.\,\ref{merger}). Numbers in parentheses refer to B[e] supergiant candidates
\citep[see][]{MK_Kraus09}. Note that the only known extragalactic B[e] stars are 
supergiants in the Magellanic Clouds.}
\label{MK_classes}
\smallskip
\begin{center}
{\small
\begin{tabular}{lllll}
\tableline
\noalign{\smallskip}
 & Herbig B[e] & cPNe B[e] & B[e] supergiant & symbiotic B[e] \\
\noalign{\smallskip}
\tableline
\noalign{\smallskip}
 B-type    & B-type pre-main & obscured O-type & B-type     & obscured hot \\
 spectrum  & sequence star   & white dwarf     & supergiant & compact obj. \\
\noalign{\smallskip}
\tableline
\noalign{\smallskip}
 forbidden & reflection     & PN nebula     & high-density  & associated \\
 emission  & nebula         &               & non-spherical & nebula     \\
 lines     &                &               & wind          &            \\
\noalign{\smallskip}
\tableline
\noalign{\smallskip}
 dust and     & pre-main        & high-density & high-density & accretion \\
 Balmer       & sequence        & dusty disk   & (outflowing ?) & disk    \\
 lines        & accretion disk  &              & disk           &         \\
\noalign{\smallskip}
\tableline
\noalign{\smallskip}
 number of  & 4/9 & 0/12 & 0/10 (LMC) & all $\sim 200$ \\
 confirmed  &     &      & 1/4(+2) (SMC) &             \\
 binaries   &     &      & 2/(14) (MW)   &             \\
\noalign{\smallskip}
\tableline
\end{tabular}
}
\end{center}
\end{table}

Despite the classification of B[e] stars according to their evolutionary phases, more 
than half of all galactic B[e] stars still remain unclassified, since the properties of
these objects do not fit into any of these classes, nor do objects of this sub-sample 
show many common characteristics besides the B[e] phenomenon itself. It has thus been 
speculated that these stars might be binaries and that the circumstellar material, 
responsible for the B[e] phenomenon seen in these stars, might have been ejected 
during phases of binary interaction \citep[see, e.g.,][]{MK_book}. The number of 
confirmed binaries is, however, rather small. A step forward to prove this 
hypothesis might therefore be a detailed search for (especially interacting) binary 
components in these objects.

\section{Search for Binarity - More Problems Than Solutions}

The search for binarity in the unclassified B[e] stars is not straightforward.
The objects are usually faint, suffering from mostly unknown but large amounts of
predominantely circumstellar extinction. In addition, their distances are far from 
being constrained, hampering proper luminosity determinations. 

Some promising tools for finding binary components came up during the past years, and 
their applicability to B[e] stars and reliability is discussed in the following.

\subsection{Spectral Disentangling}

Spectral disentangling is a powerful tool for the analysis of composite spectra of  
eclipsing binary or multiple stars. However, the requirement for proper disentangling 
is that high-quality spectra are available and cover as much as possible of 
the orbital period \citep[see][]{MK_Hensberge}. 

In extremely rare cases composite spectra of B[e] stars have been obtained as in the
case of MWC\,623 (\citeauthor{MK_Zickgraf} \citeyear{MK_Zickgraf}; Polster et al., 
this volume), which might be a binary with a rather large period ($> 14$\,years). 
However, in general the method of spectral 
disentangling is not advisable for finding binary components in B[e] stars, because 
these stars are embedded in irregularly shaped and highly dynamical circumstellar
material, contributing and polluting significantly the photospheric lines. In addition,
several B[e] stars have been found to have dynamical atmospheres as well, resulting 
in irregular line variations, since lines of different elements are formed at 
different atmospheric depths, delivering a variety of radial velocities 
\citep[see, e.g.,][]{MK_Borges}. If such kind of lines are further contaminated by 
emission from dynamical and complex circumstellar material, the method of spectral 
disentangling can thus most probably be considered as delivering results, which might 
be rather unsatisfactory or even incorrect.

\subsection{Spectroastrometry}

A more sophisticated technique in finding especially close binaries uses 
spectro-astrometry \citep{MK_Bailey,MK_Porter}. This method has been 
successfully applied to Herbig Ae/Be stars, since it is particularly suited for 
emission line stars \citep[see, e.g.,][]{MK_Baines}. The great advantage of 
spectro-astrometry lies in its ability to detect binary companions that are fainter 
by up to 6 mag. However, the method is limited to separations larger than 100\,mas. 

The sample of \citet{MK_Baines} contained three unclassified B[e] stars: HD\,45677, 
HD\,50138, and HD\,87643. The first two were claimed to be new binary discoveries,
while the last one showed a highly complex spectro-astrometry, which, however, was
not in agreement with a binary component. HD\,87643 was thus discarded by
\citet{MK_Baines} as a binary system.

\subsection{Interferometry and Image Reconstruction}

Another tool to discover even closer binaries with separations $\ll 100$\,mas is 
provided by optical long baseline interferometry. For instance, observing classical Be 
stars, \citet{MK_Meilland} discovered a companion to $\delta$ Cen at a separation of 
68.7\,mas. Long baseline interferometry, nowadays limited to bright stars, 
additionally allows to resolve small scale structures, constraining the 
geometry of the circumstellar or circumbinary material (see, e.g., 
\citeauthor{MK_Martin} \citeyear{MK_Martin}; Bonneau, this volume).

In addition, \citet{MK_Millour} recently applied a new technique of image 
reconstruction for the VLTI/AMBER observations of the B[e] star HD\,87643,
based on which they resolved a cool companion at a separation of 34\,mas. This 
method thus seems to be the most powerful one to disentangle
companions at angular separation typically larger than 1\,mas, i.e. 1-2\,AU
at 1-2\,kpc, with typical periods larger than a few months. Such systems are
difficult to detect by classical spectroscopy.

\section{Binary Mergers}\label{merger}

Table\,\ref{MK_classes} highlights that all classified B[e] stars possess a 
high-density dusty disk. This feature is especially puzzling for compact PNe and 
luminous supergiants. In addition, detailed investigations on the compact PNe 
Hen\,2-90 revealed a latitude dependent ionization structure of its wind 
\citep{MK_Kraus05}, while no evidence for binarity was found \citep{MK_Kraus07}. 
Such wind structures are generally observed from rapidly rotating stars. Also, 
several Magellanic Cloud B[e] supergiants were found to be rapidly rotating 
\citep[see, e.g.,][]{MK_Kraus08}. Since rapid rotation is not 
expected for stars in these evolutionary phases, the stars must have been spun up.
The most natural cause for a star to spin up is provided by binary merger processes
\citep[e.g.,][]{MK_Podsi}. Such a scenario seems to be quite plausible, given that
about 10\,\% of all stars are expected to experience a merger process, as suggested by 
population synthesis models. For compact PNe and supergiants the B[e] phenomenon
might therefore be strictly linked to binary merger processes, though a proof of
this hypothesis, e.g., in terms of abundance studies, has not been made yet.

\section{Conclusions}

The B[e] phenomenon in the yet unclassified stars has been proposed to be due to
binary interaction. However, to test this hypothesis is a challenging task, since 
the methods of binary identification quickly reach their limit of applicability for 
stars being surrounded by highly complex and highly dynamical material.
Therefore, it is quite sobering, how little results have been achieved so far. 
Using spectro-astrometry two binary systems were detected. The distribution of 
circumstellar material of a third B[e] star studied was too complex and the 
separation of its binary component too small to be catched by spectro-astrometry. 
Instead, this system was resolved by interferometry and image reconstruction 
techniques. 

The discovery of close binaries amongst the unclassified B[e] stars is 
the first but most important step towards our understanding of their nature. The
search for indications that interaction is or has been taking place is
the next step to test the interacting binary hypothesis. This task is 
challenging, especially with respect to the large number of unclassified B[e] stars,
but with the advent of the described new techniques our odds are improving.

\acknowledgements 
M.K. acknowledges financial support from GA\,AV \v{C}R grant number KJB300030701 and 
M.B.F. from the Programme National de Physique Stellaire (France) and 
CNRS-France for the post-doctoral grant.


\end{document}